\documentclass[preprint]{revtex4}
\nonstopmode
\usepackage{amsmath}
\usepackage{amsbsy}
\usepackage{pstricks,graphicx}
\newcommand{\dirint}[3]{\ensuremath{\langle #1|#2|#3\rangle}}
\newcommand{\bs}{\boldsymbol}

\begin{document}

\title{Runge-Gross
action-integral functional
re-examined}
\author{J. Schirmer}
\affiliation{Theoretische Chemie,\\ Physikalisch-Chemisches Institut,
Universit\"{a}t  Heidelberg,\\
D-69120 Heidelberg, Germany}

\begin{abstract}
The density-based action-integral functional introduced
by Runge and Gross [Phys. Rev. Lett. 52, 997(1984)] in their foundation of 
time-dependent density-functional theory (TDDFT) is re-examined.
Based on an obvious expansion of the original definition, it becomes apparent
that the action-integral functional is both trivial and non-stationary.
It cannot be used to establish equations of motion for the time-evolution of
quantum systems at the density-function level.

\end{abstract}

\date{\today}
\maketitle

In the following, we briefly address the action-integral functional (AIF) introduced by Runge and 
Gross (RG) in their attempt to establish time-dependent density-functional theory (TDDFT)~\cite{run84:997}. 
The RG-AIF has been criticized as lacking definitness due to a purely time-dependent phase
function entering the definition of the density-based AIF~\cite{lee01:1969,sch07:022513}.
However, the phase problem may not be the main issue here. The AIF can readily be written
in a more explicit form showing that it is not stationary and does not establish  
an equation of motion at the density level. While the finding 
discussed below is entirely obvious, it seems to have escaped due attention previously. 

Let us consider an $N$-electron system subject to a time-dependent one-particle potential, 
the Hamiltonian being of the form
\begin{equation}
 \hat H(t) = \hat T + \hat V + \hat U(t)
\end{equation}
where $\hat T$ and $\hat V$ are the kinetic energy operator and 
Coulomb repulsion operator,
respectively, while 
\begin{equation}
\hat U(t) = \sum_i^N u(\bs r_i,t) 
\end{equation}
denotes the time-dependent local one-body potential operator,
comprising a static and a time-dependent part according to $\hat U(t) = \hat W + \hat F(t)$.
For definitness, we will suppose that the time-dependent ``external'' potential 
sets in at $t=0$, that is, $\hat F(t) = 0$ for $t<0$, and the system is in the ground state
$\Psi_0$
of the time-independent Hamiltonian $\hat H = \hat T + \hat V + \hat W$ at $t=0$. 
The solution $\Psi(t)$ of the time-dependent Schr\"{o}dinger equation (TDSE)
for the initial value $\Psi(0) = \Psi_0$ gives rise to the time-dependent
density function $n_0(\bs r, t)$. TDDFT claims that it is possible to
determine $n_0(\bs r, t)$ without recourse to the TDSE.

The basic entity in the original RG foundation of TDDFT is the 
density-based AIF for the system under consideration, given by 
\begin{equation}
\label{eq:aif}
A[n] = \int_{t_1}^{t_2} dt\, \dirint{\Psi[n](t)}{i\frac{\partial}{\partial t}
 - \hat{H}(t)}{\Psi[n](t)}
\end{equation}
Here $\Psi[n](t)$ is the wave function associated with the 
time-dependent density function $n(\bs{r},t)$ according to the  
first Runge-Gross (RG1) theorem~\cite{run84:997}. 
Let us recall that the RG1 theorem establishes a mapping
between time-dependent densities, $n(\bs r,t)$, and 
time-dependent ``external'' potentials, $v_{ext}[n](\bs r,t)$,
\begin{equation}
n(\bs r,t) \rightarrow v_{ext}[n](\bs r,t) + c(t)
\end{equation}
such that the solution $\Psi[n](t)$ of the $N$-electron TDSE
\begin{equation}
\label{eq:psin}
i \frac{\partial}{\partial t} \Psi[n](t)  = 
\{ \hat{T} + \hat{V} + \hat{V}_{ext}[n](t) + C(t)\}
                                \Psi[n](t)
\end{equation}
reproduces the respective density $n(\bs{r},t)$. 
Here $\hat{V}_{ext}[n](t)$ is the $N$-electron form of the external potential,
that is,
\begin{equation}
\hat{V}_{ext}[n](t) = \sum_i^N v_{ext}[n](\bs r_i,t) 
\end{equation}
Note that $v_{ext}[n](\bs r,t)$ is determined by the density only up to a
time-dependent function $c(t)$, that is, $C(t) = N c(t)$ in Eq.~(\ref{eq:psin}).
To specify the initial value problem, we may consider densities
where $n(\bs r,0) = n_0(\bs r)$, and, moreover, suppose $\Psi[n](0) = \Psi_0$.

Using that $\Psi[n](t)$ fulfills the TDSE (\ref{eq:psin}),
the RG-AIF according to Eq.~(\ref{eq:aif}) can be written 
in a simple form with an explicit Lagrange-type function,
\begin{equation}
\label{eq:aifsf}
A[n] = \int_{t_1}^{t_2} dt \int 
\left\{v_{ext}[n](\bs r,t) - u(\bs r,t)\right \} n(\bs r,t) d\bs r 
+ \int_{t_1}^{t_2} C(t)dt
\end{equation}
Here, the kinetic and Coulomb energy contributions have cancelled along with the
time derivative, and the
remaining potential energy expectation values can be expressed entirely in terms of 
the density functions $n(\bs r,t)$, since both $\hat{V}_{ext}[n](t)$ and
$\hat U(t)$ are local one-particle operators. This shows that the AIF can be defined 
directly at the level of the density functions, and its relation to the
solution of TDSE, as implied by Eqs.~(\ref{eq:aif}, \ref{eq:psin}), is an illusion.

Let us note that the indefiniteness of the AIF due to the 
$\int C(t)dt$ term has been addressed previously (see Refs.~\cite{lee01:1969,sch07:022513,sch10:052510}), and we shall ignore it in the present 
context by supposing $C(t)=0$ in Eq.~(\ref{eq:aifsf}).

Obviously, $A[n]$ vanishes for $n(t) = n_0(t)$ since
\begin{equation}
\label{eq:vextu}
v_{ext}[n_0](\bs r,t) = u(\bs r,t)
\end{equation} 
which simply reflects the construction underlying $v_{ext}[n](\bs r,t)$. Eq.~(\ref{eq:vextu})
is not an equation of motion, nor can it be seen as a realistic means to 
determine $n_0(t)$.
An eventual solution would require guessing the potential-functional 
$v_{ext}[n](\bs r,t)$ (or an approximation to it) and solving the implicit equation (\ref{eq:vextu}), possibly by adopting a fixed-point iteration scheme.

Even more disturbing is the observation that $A[n]$ is not stationary at $n_0(t)$. 
This can be seen by evaluating the variation according to Eq.~(\ref{eq:aifsf}) and using
Eq.~(\ref{eq:vextu}):
\begin{equation}
\left .\delta A[n]\right |_{n_0} = \int_{t_1}^{t_2} dt \int d\bs r\, n_0(\bs r,t) \left.\delta v_{ext}[n](\bs r,t)\right |_{n_0}
\end{equation}
Here one cannot expect that $\left. \delta v_{ext}[n](\bs r,t)\right |_{n_0}$ vanishes, since
$n_0(\bs r,t)$ is just an ordinary density-function argument for the external potential-functional,
not distinguished from other densities.

Assuming for simplicity that $v_{ext}[n](\bs r,t)$ will depend only on $n(\bs r,t)$ (and not
on the first and higher time derivatives of $n(\bs r,t)$), the functional derivative 
of $A[n]$ takes on the form
\begin{equation}
\label{eq:aifxy}
\frac{\delta A[n]}{\delta n(\bs r,t)} = v_{ext}[n](\bs r,t) - u(\bs r,t)
+ n(\bs r,t) \frac{\partial v_{ext}[n]}{\partial n(\bs r,t)} 
\end{equation}
which shows, according to
\begin{equation}
\left. \frac{\delta A[n]}{\delta n(\bs r,t)}\right |_{n_0} = 
n_0(\bs r,t) \left.\frac{\partial v_{ext}[n]}{\partial n(\bs r,t)}\right |_{n_0} \neq 0 
\end{equation}
that the functional derivative does not vanish at the desired density $n_0(t)$ .
This means that Eq.~(11) in Ref.~\cite{run84:997} is patently wrong.

In conclusion, 
the RG-AIF can readily be written in a more explicit form, which 
makes apparent that
\begin{itemize}
 \item[(i)]
it is an essentially trivial construct that does not establish an equation 
of motion for the time-evolution of a quantum system at the density level;
\item[(ii)]
it is not stationary for the density of the system under consideration.
\end{itemize}

It should be noted that the original RG foundation of TDDFT, based on the 
stationarity of the RG-AIF, was abandoned by its principal architects some time ago,
notwithstanding
occasional attempts at a rehabilitation (see, e.g. Ref.~\cite{vig08:062511}).
The alternative offered to establish time-dependent Kohn-Sham (TDKS) equations
is based entirely on the RG1 mapping theorem (see Ref.~\cite{mar04:427}). 
Unfortunately,
the mapping foundation has not been fully disclosed in the TDDFT literature so far.
For a discussion of the problems arising here, the reader is referred to 
Refs.~\cite{sch07:022513,sch08:056502}.

\acknowledgements 
The author thanks H.-D. Meyer, L.S. Cederbaum, A. Dutoi, and A. Dreuw for clarifying discussions.

%\bibliography{DFT,Books}

\end{document}